\documentclass[useAMS,usenatbib]{mn2e}
\pdfoutput=1
\usepackage{graphicx}

\begin{document}

\title[The AMR in the SMC periphery]{The age-metallicity relationship in the Small
Magellanic Cloud periphery}

\author[A. E. Piatti]{Andr\'es E. Piatti$^{1,2}$\thanks{E-mail: 
andres@oac.uncor.edu}\\
$^1$Observatorio Astron\'omico, Universidad Nacional de C\'ordoba, Laprida 854, 5000, 
C\'ordoba, Argentina\\
$^2$Consejo Nacional de Investigaciones Cient\'{\i}ficas y T\'ecnicas, Av. Rivadavia 1917, 
C1033AAJ, Buenos Aires, Argentina \\
}

\maketitle

\begin{abstract} 
We present results from  Washington $CT_1$ photometry for eleven
star fields located in the western outskirts of the Small Magellanic Cloud (SMC), which 
cover angular distances  to its centre from 2 up to 13 degrees ($\approx$ 2.2 $-$ 13.8 kpc). 
The colour-magnitude diagrams, cleaned from the unavoidable Milky Way (MW) and background
galaxy signatures, reveal
that the most distant dominant main sequence (MS) stellar populations from the SMC centre are 
located at an angular distance of $\sim$ 5.7 deg (6.1 kpc); no sign of farther 
clear SMC MS
is visible other than the residuals from the MW/background field contamination.
The derived ages and metallicities for the dominant stellar populations of the western SMC 
periphery show a constant metallicity 
level ([Fe/H]  = -1.0 dex) and  an approximately constant 
age value ($\approx$ 7$-$8 Gyr). Their age-metallicity relationship (AMR)  do not clearly differ
 from the
most comprehensive AMRs derived for almost the entire SMC main body.
Finally, the range of ages of the dominant stellar populations in the western SMC periphery 
confirms that the major stellar mass formation activity at the very early galaxy epoch peaked
$\sim$ 7-8 Gyr ago.
\end{abstract}

\begin{keywords}
techniques: photometric -- galaxies: individual: SMC -- Magellanic
Clouds
\end{keywords}

\section{Introduction}

Our knowledge about the age-metallicity relationship (AMR) of the Small Magellanic Cloud (SMC)
has improved since recent -some of them still ongoing- systematic photometric studies have been
carried out 
\citep[e.g.][]{ss12,cetal13b,wetal13,retal15}. From an overall consensus about the formation and 
chemical evolution of this galaxy, some features can be highlighted: i) an initial period of low 
star formation rate (SFR); ii) an older enhanced star formation process that peaked at $\sim$ 
8 Gyr ago; iii) another periods of intense SFR at $\sim$ 1.5$-$2.5 and 5$-$6 Gyr, consistent with 
suggestions of close encounter with the Large Magellanic Cloud (LMC) and/or the Milky Way (MW); 
iv) field stars do not possess gradients in age and metallicity;
 v) an AMR with a
clear steady enrichment of the galactic metal content with time (older stellar populations
are more metal-poor), with some variations, among others.

These main results picture the presently known assembled scenario for the formation and chemical 
evolution of the SMC main body; the farthest studied star field is located at $\sim$ 6.5 kpc from 
its centre \citep{ng07,cetal08,netal09}. However, the SMC periphery, particularly the portion 
towards the opposite direction to the LMC, could harbour primordial stellar populations 
not yet comprehensively dealt with. \citet{netal11} carried out a photometric survey of the SMC 
periphery, from which they extracted a modelled galactic geometry using giant star counts; no detailed 
study about the spatial behaviour of star field ages and metallicities were performed.

\citet[][hereafter P12]{p12a} accomplished an AMR study of the SMC stellar populations from
 Washington $CT_1$ photometry of 12 star fields (36$\times$36 armnin$^{\rm 2}$ each) getting a compromise 
between covering a relatively large area and reaching the oldest main sequence turnoffs (MSTOs). He 
found that the field stars do not possess gradients in age and metallicity, and that stellar populations 
formed since $\sim$ 2 Gyr ago are more metal-rich than [Fe/H] $\sim$ -0.8 dex and are confined to
the innermost regions (semi-major axis $\le$ 1 deg), along with relatively more metal-poor 
([Fe/H] $\sim$ -1.0 $-$ -1.5 dex) and older (age $\sim$ 3$-$8 Gyr) field stars.
He also compared the field star AMR
to that of the star cluster population with ages and metallicities in the same star field scale, and
found that clusters and star fields share similar chemical evolution histories. Nevertheless, the
outermost star field studied by P12 is located $\sim$ 4.9 kpc, so that the AMR in the SMC periphery
has remained mostly unstudied.

In this paper we continue the work of P12 by studying other 11 SMC fields located in the western outskirts 
of the SMC. As a matter of consistency, we also analyse a  Washington $CT_1$ data set using the same
techniques. We aim at characterizing the dominant stellar populations distributed in that region
and providing for the first time with a representative AMR. For these purposes, we structured the
paper as follows: Section 2 deals with the description of the photometric data sets, the reduction
processes and the various calibrations involved, including error and completeness analyses. The
observed colour-magnitude diagram (CMD) features as well as the cleaning procedures for decontaminating 
them from
the MW/background galaxy signatures are described in Section 3. In Section 4 we estimate the 
ages and metallicities
of the dominant stellar populations in each SMC field, and discuss their AMR to the light of the
most comprehensively known SMC AMRs. Concluding remarks are given in Section 5.

\section{Data handling}

In our previous series of studies of SMC clusters and star fields \citep[see, e.g.][P12]{p11b} we 
used the Washington  $CT_1$ photometric system \citep{c76} from which ages and metallicities for stellar 
populations older than $\sim$ 1 Gyr can be estimated \cite[and references therein]{pp15}. For 
these reasons, and in order to keep consistency 
with our previous studies, we performed a search within the National Optical Astronomy Observatory 
(NOAO) Science Data Management (SDM) Archives\footnote{http://www.noao.edu/sdm/archives.php.} seeking 
for Washington photometric data towards the SMC periphery. As a result, we found images 
obtained at the Cerro-Tololo Inter-American Observatory (CTIO) 4m Blanco telescope with the 
Mosaic\,II camera 
(36$\times$36 arcmin$^{\rm 2}$ field with a 8K$\times$8K CCD detector array, scale 0.274 arcsec/pixel) 
covering 11 fields in the SMC outskirts for a total 
area of $\sim$ 4.0 square degrees (programme CTIO 2006B-0013, PI: Saha).
The log of the observations
is presented in Table 1, where the main astrometric, photometric and observational information is
 summarized. Note that the data set includes the filter $R$ which is a more efficient substitute of
the $T_1$ filter \citep{g96}, but our final standard magnitude is the Washington $T_1$ mag. Fig. 1 
depicts the spatial distribution of the SMC fields, where we included those
in P12 for comparison purposes. The maximum angular distance probed from the centre of 
the SMC is $\sim$ 13.0 deg. 

The  Washington $CT_1$ photometric data set used in this work were obtained from a comprehensive 
process that involved the reduction of the raw images, the determination of the instrumental photometric
magnitudes, and the standardization of the photometry. We have already described in detail such steps not 
only in the works cited above, but also in \citet{p11a,p11b,p11c,p12c,pb12,pietal12,metal14}. For this reason, 
we summarize here some specific issues in order to provide the reader with an overview
of the photometry quality. The data reduction followed the procedures documented by the NOAO Deep Wide 
Field Survey team \citep{jetal03}
and utilized the {\sc mscred} package in IRAF\footnote{IRAF is distributed by the National 
Optical Astronomy Observatories, which is operated by the Association of 
Universities for Research in Astronomy, Inc., under contract with the National 
Science Foundation.}. We performed overscan, trimming and cross-talk corrections, bias subtraction,
flattened all data images, etc., once the calibration frames (zeros, sky- and dome- flats, etc) were properly 
combined. For each image we obtained
 an updated world coordinate system (WCS) database with a rms error in right ascension and 
declination smaller than 0.4 arcsec, by using $\sim$ 500 stars catalogued by the 
USNO\footnote{http://www.usno.navy.mil/USNO/astrometry/optical-IR-prod/icas/usno-icas}. 

We also retrieved MOSAIC\,II and CTIO 0.9m CFCCD images of the standard fields 
SA\,92 and SA\,98 observed as part of the same programme. 
The reduction 
of the MOSAIC\,II data set was performed following the steps as for the SMC fields.
These standard fields contain
between 8 and 14 standard stars each \citep{g96} and series of eight observations were carried out by 
dithering the images, 
so that most of the standard stars were distributed on each of the MOSAIC\,II's eight chips. 
This observing procedure was repeated for different airmasses. 
The images from the CTIO 0.9m telescope were taken with
the Tektronix 2K\#3 CCD, using quad-amp readout.  The scale on the chip is 0.4 arcsec/pixel
yielding an area covered by a frame of 13.5$\times$13.5 arcmin$^{\rm 2}$. The integrated IRAF-Arcon 3.3 interface for 
direct imaging was employed as the data acquisition system. 
The data were processed using the {\sc QUADPROC} package in IRAF.
After applying the overscan-bias 
subtraction for the four amplifiers independently, we carried out flat-field 
corrections using a combined sky-flat frame, which was previously checked for 
a non-uniform illumination pattern with the averaged dome-flat frame.

We derived nearly 80 independent magnitude measures of standard stars 
per filter for each night using the {\sc apphot} task within IRAF, in order to secure the
 transformation from the 
instrumental to the Washington  $CT_1$ standard system.  The relationships between instrumental and 
standard magnitudes were obtained by 
fitting the equations:

\begin{equation}
c = a_1 + T_1 + (C-T_1) + a_2\times X_C + a_3\times (C-T_1),
\end{equation}

\begin{equation}
r = b_1 + T_1 + b_2\times X_R + b_3\times (C-T_1),
\end{equation}

\noindent where $a_i$ and $b_i$ ($i$ = 1, 2, and 3) are the fitted coefficients, and $X$ 
represents the effective
airmass. Capital and lowercase letters represent standard and instrumental
magnitudes, respectively. 
 We solved the transformation equations with the {\sc fitparams} task in IRAF for
each night, and substituted the derived mean zero point and colour term  
($a_3$ = -0.090, $b_3$ = -0.020) values for each observing run 
into the above equations and solved  for the airmass coefficient for each night. 
Typical values of $a_2$ and $b_2$ were 0.305 and 0.095 mag/airmass, respectively.
The nightly rms errors from 
the transformation to the  standard system were 0.021 and 0.017 mag for $C$ and $T_1$,
 respectively, 
indicating the nights were of excellent photometric quality. 

The stellar photometry was performed using the star-finding and point-spread-function (PSF) fitting 
routines in the {\sc daophot/allstar} suite of programs \citep{setal90}. We measured magnitudes on 
the single 
image created by joining all 8 chips together using the updated WCS. This allowed us to use a unique 
reference coordinate system
for each SMC field. For each Mosaic image, a quadratically varying 
PSF was derived by fitting $\sim$ 1000 stars (nearly 110-130 stars per chip), once the neighbours
 were eliminated using a preliminary PSF
derived from the brightest, least contaminated $\sim$ 250 stars (nearly 30-40 stars per chip). Both
 groups of PSF 
stars were interactively selected. We then used the {\sc allstar} program to apply the resulting
 PSF to the 
identified stellar objects and to create a subtracted image which was used to find and measure
 magnitudes of 
additional fainter stars. This procedure was repeated three times for each frame. 
We computed aperture corrections from the 
comparison of PSF and
aperture magnitudes by using the neighbour-subtracted PSF star sample.
After deriving the photometry for all detected objects in the $C$ and $T_1$ filters,
a cut was made on the basis of the parameters
returned by DAOPHOT. Only objects with $\chi$ $<$2, photometric error less than 2$\sigma$ above
 the mean error at a given magnitude, and $|$SHARP$|$ $<$ 0.5 were kept in each filter, and then 
the remaining objects in the $C$ and $T_1$ lists 
matched with a tolerance of 1 pixel. 

Finally, we standardized
the resulting instrumental magnitudes and combined all the independent measurements using the
 stand-alone {\sc daomatch} 
and {\sc daomaster} programs\footnote{Provided kindly by Peter Stetson.}.  The final information for 
each field 
consists of a running number per star, its right ascension and declination coordinates, the averaged 
$T_1$ magnitude and $C-T_1$ colour, the standard errors $\sigma(T_1)$ and $\sigma(C-T_1)$ and
the number of measurements in the $C$ and $T_1$ filters. Whenever a star does have only one measure of $C$ and $T_1$
mags, the errors provided by the {\sc daophot/allstar} routines were listed.  
Table 2 gives this information for Field\,1. Only a portion 
of this table is shown here for guidance regarding its form and content. The whole content of  Table 2, as well as
the final information for the remaining fields, is 
available in the online version of the journal.

As is well known, photometric errors, crowding effects and the detection limit
of the images cause incompleteness and therefore results in the increasing loss of stars at fainter
 magnitudes. 
Commonly, artificial star tests on the deepest images are performed in order to derive the 
completeness
level at different magnitudes. 
We used the stand-alone {\sc addstar} program in the {\sc daophot}
package \citep{setal90} to add synthetic stars,
generated at random with respect to position and magnitude, to each deepest image in order to derive its
completeness level. We added a number
of stars equivalent to $\sim$ 5$\%$ of the measured stars in order to avoid in the synthetic images 
significantly 
more crowding than in the original images. On the other hand, to avoid small number statistics in the
 artificial-star 
analysis, 
we created five different images for each original one. We used the option of entering the number of
 photons
per ADU in order to properly add the Poisson noise to the star images. 

We then repeated the same steps to obtain the photometry of the synthetic images as described above, 
i.e., 
performing three passes with the {\sc daophot/allstar} routines. 
The star-finding efficiency were estimated by comparing the output 
and the input data for these stars using the {\sc daomatch} and {\sc daomaster} tasks.
We illustrate in Fig. 2 the resultant completeness fractions in the $T_1$ versus 
$C-T_1$ CMD for the most pupolated field in our
sample (Field\,11). Fig. 2 shows that the 50$\%$ completeness level is located at
 $C$ $\sim$ 23.75 mag and $T_1$ $\sim$ 24.25 mag; for less crowded fields this
completeness level reaches down to $\sim$ 0.75-1.00 mag fainter, depending on the
crowding. Note that, according to the theoretical isochrones computed by
\citet{betal12} and considering the SMC distance modulus $(m-M)_o$ = 18.96
\citep{dgb15}, a 13 Gyr old stellar population ([Fe/H] $\approx$ -1.3 dex) should 
have its MSTO at $T_1$ $\approx$ 22.5 mag. Therefore, our photometry reaches the 
50$\%$ completeness level 2-3 mag below the oldest known SMC MSTOs \citep{cetal08,netal09}.

\section{Global properties of the colour-magnitude diagrams}

In Fig. 3 we plot the CMDs for all measured stars in each SMC field. We measured from 
$\sim$ 11 up to 76 thousand stars from the outermost to the innermost fields, with
an average of $\sim$ 33 thousand stars, thus yielding a $CT_1$ photometric database 
of $\sim$  175 thousand stars. The CMDs present a strong signature of the  foreground/background
fields which blurs the SMC outskirts main features. Nevertheless,
it can be seen an old main sequence (MS) population in Field\,7 ($C-T_1$ $\approx$ 0.8, 
$T_1$ $\sim$ 22-23), and increasing younger populations (brighter $T_1$ MSTOs) for inner 
SMC fields. Thus, the series of panels depicted in Fig. 3 show the age transition of the 
stellar populations in the outermost SMC regions. At the right margin of each panel we show the
behaviour of the errors $\sigma$($T_1$) and $\sigma$($C-T_1$). 
As can be seen, it would seem that a relatively small
dispersion accompanies the intrinsic features along almost $\sim$ 9 mags.

\subsection{Cleaning the CMDs}

In order to disentangle the MW/background signature  
in the peripheral SMC CMDs, we applied
a procedure primarily developed by \citet[][see, Fig. 12]{pb12} to clean 
star cluster CMDs from field star contamination \citep[see, e.g.][]{p14,petal14b,petal15a}. 
Comparisons of field and cluster CMDs have long been done by comparing
the numbers of stars counted in boxes distributed in a similar manner
throughout both CMDs. However, since some parts of the CMD are more
densely populated than others, counting the numbers of stars within
boxes of a fixed size is not universally efficient. For instance, to
deal with stochastic effects at relatively bright magnitudes (e.g.,
fluctuations in the numbers of bright stars), larger boxes are
required, while populous CMD regions can be characterized using
smaller boxes. Thus, use of boxes of different sizes distributed in
the same manner throughout both CMDs leads to a more meaningful
comparison of the numbers of stars in different CMD regions. 

For our purposes, we adopted Field\,1 as the  MW/background field CMD, which
is located $\sim$ 3$-$4 deg farther than the SMC boundary delimited by \citet{netal11}. 
We also assumed that the MW/background field is uniform in terms of luminosity
function, colour distribution and stellar density  across the
surveyed region \citep{betal15}. Note that the eleven SMC fields are projected over a 
relatively small MW area of $\sim$ 16$\times$7 deg$^{\rm 2}$.
We checked this assumption for the MW by building a series of synthetic CMDs
using the Besan\c{c}on galactic model \citep{retal03}.  Note that MW models can only 
include MW stars and not distant unresolved galaxies, which occur in significant numbers at 
faint magnitudes. The typical number of generated stars
represent $\sim$ 50 per cent of the stars measured in Field\,1.
Consequently, the present cleaning procedure allows us to recognize 
the location of the bulk of the SMC population, while more sophisticated approaches 
might provide more quantitative outcomes \citep[e.g.][]{roderick15}.

We then compares the MW/background CMD --previously corrected by the difference in reddening 
(see column 2 of Table 3) and in completeness effects -- to each 
one of the SMC CMDs
and subtract from the latter a representative  MW/background CMD in terms of stellar density, 
luminosity function and colour distribution. The representative 
 MW/background CMD was built from scanning the observed  MW/background CMD using 
boxes that vary in size, thus
achieving a better  MW/background representation than counting the number of stars in boxes fixed 
in size. Then, the stars in the SMC CMDs that fall within the defined boxes and
closest to the representative positions were eliminated. 

The method allows that boxes vary in magnitude and colour separately according to the
free path between stars in the CMD, so that they result bigger in CMD regions with a 
small number of stars, and vice versa.  The free path is defined as
$\Delta$(colour)$^2$ + $\Delta$(magnitude)$^2$  = (free path)$^2$, 
where $\Delta$(colour) and $\Delta$(magnitude) are the distances from the considered star to the 
closest one in abscissa and ordinate in the  MW/background CMD. In practice, an initial 
reasonably large box with a dimension of ($\Delta$(colour), $\Delta$(magnitude) = (0.5, 1.0) is 
centred on each single star in the  MW/background CMD. Then, the method 
subsequently reduces the box sizes until they reach the stars closest in magnitude and colour, separately, 
so that the resulting closest magnitude and colour
will be used to define the free path of the considered star. The task is 
repeated for every star in the  MW/background CMD. Therefore, each star in the  MW/background 
CMD has associated a different box. 

These boxes are then superimposed to the SMC CMDs, and the stars closest to their centres
are eliminated. In this sense, the box sizes depend not 
only on the stellar density of the MW/background field (the denser a field the more stars 
in the MW CMD), 
but also on the magnitude and colour distributions
of those stars in the MW/background CMD, making some parts of the MW/background CMD  
more populated than 
others. For instance, relatively bright field red giants with small photometric errors
usually appear relatively isolated at the top-right zone of the CMD, while faint
MS stars are more numerous at the bottom part of the CMD. For this reason, 
bigger boxes are required to satisfactorily subtract stars from the SMC CMD regions
where there is a small number of  MW/background stars, while smaller boxes are necessary for 
those CMD regions more populated by  MW/background stars.

\subsection{Hess diagrams}

Hess diagrams have played an increasingly prominent role in the analysis of 
photometric data, in particular of the Magellanic Clouds. They show the frequency or density 
of occurrence of stars at various positions on the 
CMD, thus providing a star density map in the CMD. Hess diagrams have been profitably exploited
by using different robust techniques \citep[][and references therein]{hz01,d02}. Here, we simply
take advantage of Hess diagrams to identify the prevailing features -in terms of density of stars- 
and their intrinsic dispersions in the cleaned SMC CMDs. In order to
produce these Hess diagrams we counted the number of stars placed in different magnitude-colour bins 
with sizes 
[$\Delta$$T_1$, $\Delta$$(C-T_1)$] = (0.1,0.05) mags and then we represented the resultant count
 scales with a 10 grey 
levels logarithmic scale. The resultant Hess diagrams for the observed and cleaned
SMC CMDs are depicted in a series of panels in Fig. 4.

The cleaned Hess diagrams exhibit different features 
which tell us about the stellar populations in the surveyed SMC region. A mixture of intermediate-age 
through old stellar populations clearly appears to be the main feature
of these Hess diagrams. Since the SMC field CMDs are obviously composed of stars of different stellar
populations, the respective Hess diagrams allow us to recognize the so-called "dominant populations"
(generally darker regions in the Hess diagrams). These stellar populations are the most numerous 
in a particular region and can be referred as the "representative" stellar population along the 
line of sight (LOS) of that SMC field \citep[P12,][]{pg13,petal14c}. The definition of a 
representative population
could not converge to any dominant population if the stars in a given field came from a constant 
SFR integrated
over all time. For our fields, however, we could clearly identify the respective most 
populated features. Particularly, the representative MSTOs are 
in average $\sim$ 1.5 mag brighter 
than the $T_1$ mags for the 100$\%$ completeness level of the respective field, i.e., 
the faintest $T_1$ mag where completeness
 is still 100$\%$. Therefore, we actually reach the MSTO of the representative population of each field
 with a negligible loss of stars at that magnitude.

The most distant MS populations from the SMC centre
detected by our photometry are located at an angular distance of $\sim$ 5.7$\degr$ (Field\,7), 
equivalent to 6.1 kpc if the SMC distance modulus of 18.96 mag is adopted \citep{dgb15}. 
The adopted optical SMC centre is : 00$^{h}$ 52$^{m}$ 45$^{s}$, $-$72$\degr$ 49$\arcmin$ 
43$\arcsec$ (J2000) \citep{cetal01}. 
We did not find any visible sign of farther  clear SMC MS other than the residuals from the 
 MW/background contamination.
The resulting SMC radius is in very good agreement with those obtained
by \citet[][$\sim$ 6.3 kpc]{ng07} from $BV$ photometry of 3 SMC fields and by 
\citet[][$\sim$ 6 kpc]{dpetal10} from Ca\,II triplet spectroscopy of giant stars, among others.
However, it results shorter than the SMC peripheral boundary estimated by 
\citet[][$\sim$ 10.7 kpc]{netal11}. Their data cover a much larger area 
of the SMC, which allowed them from the analysis of the giant stellar component to find evidence
for a break population beyond $\sim$ 8 kpc with a shallow radial density profile 
that could be either a bound stellar halo or a population of extratidal stars.
In this context,
our results
confirm that the bulk of the SMC population is contained within $\sim$ 7 deg from
the galaxy centre, independently of any possible dependence of the SMC extent 
with the position angle, beyond which there is a low density envelope extending
further out found by \citep{netal11}. Note additionally that some previous studies have suggested the existence of different asymmetries 
in the SMC 
\citep[e.g.][]{ss12,cetal13b,retal15}.  Nevertheless, we also carried out counts of red giant 
clump stars within the box ($\Delta$($C-T_1$),$\Delta$($T_1$)) = (0.95$-$1.70, 17.90$-$19.70)
in the cleaned and observed CMDs. The results are depicted in Fig. 5, where we added a constant to the
counts in the cleaned CMDs for comparison purposes and indicated  the
count level for the outermost field Field\,1 (at 12.8 deg, equivalent to 13.8 kpc) with an horizontal line.
The uncertainties introduced by both the errors in the relative SMC field reddening and in the photometry
are much smaller than the symbol size.
The similar shape of both profiles tell us about the robustness of the cleaning procedure, since
a similar number of stars was subtracted from each field, once they were properly corrected
by incompleteness and reddening effects. On the other hand, the profiles  show
that the SMC red clump star population  that our analysis is able to detect  
would not appear to extend beyond $\sim$ 6$-$7 deg ($\sim$ 6.6$-$7.7 kpc).

\section{Analysis}

We used the cleaned CMDs and Hess diagrams of Fields\,7 to 11
to constrain the fundamental SMC field parameters by matching the observations with the theoretical 
isochrones of \citet{betal12} and by using the new age$-$metallicity diagnostic diagram
for the Washington photometric system \citep{pp15}. 

The estimation of the mean SMC field reddening values was made by taking advantage of the NASA/IPAC 
Extragalactic Data base\footnote{http://ned.ipac.caltech.edu/. NED is operated by the Jet Propulsion 
Laboratory, California Institute of Technology, under contract with NASA.} (NED) to obtained Galactic 
foreground reddening values for the SMC field list. 
We adopted the recent recommendation for the mean SMC distance modulus $(m-M)_0$ = 18.96 mag, whose 
formal uncertainty is 0.02 mag but can reach 0.15-0.20 mag if the complex SMC geometry is particularly 
considered \citep{dgb15}. 
Bearing in mind a 1$\sigma$ LOS depth of 6 kpc \citet{debetal15}, we found that the difference in 
apparent distance moduli could be as large as $\Delta(T_1 - M_{T_1})$ $\sim$ 0.2 mag. This difference is
smaller than the difference between two closely spaced isochrones as used here ($\Delta$ log($t$ yr$^{-1}$) 
= 0.1, $\Delta$($M_{T_1}$ at the MSTO-subgiant branch region) $\sim$ 0.3 mag), so that adoption of a unique 
value for the distance modulus does not dominate the final error budget incurred in matching isochrones to 
the SMC field Hess diagrams.

We used the ridge lines of the representative SMC field MSs in the Hess diagrams (darker places)
to properly find the theoretical isochrones which best superimpose on them. As for the metallicity, we 
used the values derived from the new age$-$metallicity diagnostic diagram  based  
on the magnitude difference between the red giant clump and the MSTO
\citep[][see their fig. 4]{pp15}.
We assigned the age of such isochrones to 
the representative  ages of each SMC field. 
Table 3 lists the derived representative ages and metallicities and their dispersions. The  age 
dispersions have been estimated bearing in mind the broadness in magnitude and colour of the dominant MSs, which 
represent in general a satisfactory estimate of the  age spread around the prevailing population. 
The metallicity dispersions come from the estimated spread in magnitude of the red clump and
the MSTO.
Fig. 6 depicts the 
cleaned Hess diagrams with three different theoretical isochrones superimposed for comparison 
purposes.


The derived representative ages and metallicities  suggest
a constant metallicity level ([Fe/H]= -1.0 dex) and  an approximately constant age
value ($\approx$ 7-8 Gyr) for the western SMC periphery. 
We plot these values in Fig. 7 with solid circles, whereas we represent their dispersions 
with errorbars. Note that the youngest age (5 Gyr) correspond to Field\,11, which is located within the 
SMC main body. Similar age/metallicity values were found by \citet{cetal08} from Ca\,II triplet
spectroscopy of giants in 4 fields located towards the western side of the SMC main body,
as well as by \citet{netal09} who studied 12 SMC star fields from $BR$ photometry
(see Fig. 7). 
The present AMR  does not clearly differ from 
that obtained by P12 for the SMC main body using the 
same techinques, reproduced by open circles in Fig. 7.
The mean SMC AMR derived by P12 is in an overall very good agreement with those obtained by 
\citet[][see their fig. 14 and discussions therein]{cetal13b}, \citet[see their fig. 17 and discussions 
therein]{retal15}, among others. 

We did not find in the western SMC periphery dominant stellar populations older than $\sim$ 10 Gyr,
but a peak centred at $\sim$ 7-8 Gyr. This peak is in excellent agreement with the SFRs recovered by 
\citet{cetal08,netal09,cetal13b} and \citet[][see their fig. 16]{retal15}, who also showed a low initial 
SFR at $\sim$ 10 Gyr.  Particularly, \citet{retal15} showed that the stellar mass formation 
-measured as the relative to the total galaxy mass ratio- reached a peak $\sim$ 8 Gyr ago. In this
context, it seems that the western SMC periphery does not distinguish itself from the major formation
processes that took place during the galaxy formation epoch.

\section{Conclusions}

We have study the western SMC periphery from Washington  $CT_1$ photometry of eleven
star fields (36$\times$36 arcmin$^{\rm 2}$ each) covering angular distances to the SMC centre
from 2 up to $\sim$ 13 degrees ($\approx$ 2.2 $-$ 13.8 kpc). Our photometric database reaches 
the 50$\%$ completeness level
$\sim$ 2$-$3 mag below the oldest SMC MSTOs, depending on the crowding.

We cleaned the noticeable  MW/background field contamination present in the observed SMC star 
field CMDs using
a procedure which makes use of boxes of variable side centred on every star in the 
MW/background CMD, in order
to tightly reproduce it in terms of stellar density, luminosity function and colour
distribution. We subtracted the stars in the SMC CMDs that fall within the defined boxes
and are closest to the defined box centres.

From the cleaned CMDs (and hence the respective Hess diagrams) we identified the representative
(most numerous) stellar populations of each field. Particularly, the most distant representative
MS populations from the SMC centre detected by our photometry are located at an angular distance
of $\sim$ 5.7 deg (6.1 kpc); no sign of farther SMC MS stars is visible other than the residuals from 
the MW contamination.

The derived representative ages and metallicities  suggest a constant metallicity level ([Fe/H] 
= -1.0 dex) and an approximately constant age ($\approx$ 7$-$8 Gyr) for the 
composite stellar population of the western SMC periphery.
These values  do not clearly differ from those of the
most comprehensive AMRs derived for almost the entire SMC main body.

Finally, the range of ages of the representative stellar populations in the western SMC periphery confirm 
previous results about the major star formation processes from which the galaxy was born. Indeed, 
after a low SFR at $\sim$ 10 Gyr, the SMC experienced its first significative star formation
activity period which peaked $\sim$ 7-8 Gyr ago.

\section*{Acknowledgements}
 I am grateful for the comments and suggestions raised by the anonymous
referee which helped improve the manuscript.
This work was partially supported by the Argentinian institutions
CONICET and Agencia Nacional de Promoci\'on Cient\'{\i}fica y
Tecnol\'ogica (ANPCyT).
This research uses services or data provided by the NOAO Science Archive. 
NOAO is operated by the Association of Universities for Research in Astronomy 
(AURA), Inc. under a cooperative agreement with the National Science Foundation.

\bibliographystyle{mn2e_new} 
\bibliography{paper} 
%
%

\clearpage

\begin{table*}
\caption{Star fields in the SMC.}
\begin{tabular}{@{}lcccccccc}\hline
Field  & $\alpha$$_{2000}$ & $\delta$$_{2000}$  &  {\it l}  &   b   &  Filter &  Exposure & Airmass & Seeing\\
    &  $^{(h m s)}$  &  ($\degr$ $\arcmin$ $\arcsec$) & ($\degr$) & ($\degr$) & (mag)    &   (s)     &     &  (arcsec) \\\hline
1  &  22  35  28 & $-$67  06  42 & 320.8 & -45.0 &  $C$ & 1$\times$60 + 1$\times$300 + 3$\times$1080 & 1.25-1.27 & 1.3-1.6 \\
   &             &               &       &       &  $R$ & 1$\times$10 + 1$\times$50 + 3$\times$580 &   1.25-1.26 & 1.3-1.6 \\
2  &  22  52  43 & $-$68  05  45 & 318.2 & -45.4 &  $C$ & 1$\times$60 + 1$\times$300 + 3$\times$1080 & 1.27-1.28 & 1.0-1.6 \\
   &             &               &       &       &  $R$ & 1$\times$10 + 1$\times$50 + 3$\times$580 &   1.28-1.30 & 1.0-1.2 \\
3  &  23  03  52 & $-$69  18  00 & 316.0 & -45.0 &  $C$ & 1$\times$60 + 1$\times$300 + 3$\times$1080 & 1.29-1.32 & 1.1-1.3 \\
   &             &               &       &       &  $R$ & 1$\times$10 + 1$\times$50 + 3$\times$580 &   1.29      & 0.9-1.0 \\
4  &  23  25  00 & $-$70  24  00 & 313.0 & -45.0 &  $C$ & 1$\times$60 + 1$\times$300 + 3$\times$1080 & 1.40-1.52 & 1.3-1.5 \\
   &             &               &       &       &  $R$ & 2$\times$580 &                               1.54-1.56 & 1.2 \\
5  &  23  32  08 & $-$70  41  27 & 312.1 & -45.0 &  $C$ & 1$\times$60  + 3$\times$1080 &               1.38-1.46 & 1.4-1.6 \\
   &             &               &       &       &  $R$ & 1$\times$10 +  3$\times$580 &                1.32 & 1.1-1.3 \\
6  &   0  04  31 & $-$66  22  33 & 310.2 & -50.1 &  $C$ & 1$\times$60 + 1$\times$300 + 3$\times$1080 & 1.24-1.25 & 1.3-1.6 \\
   &             &               &       &       &  $R$ & 1$\times$10 + 1$\times$50 + 3$\times$580 &   1.24-1.26 & 1.0-1.2 \\
7  &  23  55  20 & $-$69  28  35 & 310.1 & -46.9 &  $C$ & 1$\times$60 + 1$\times$300 + 3$\times$1080 & 1.43-1.60 & 1.0-1.2 \\
   &             &               &       &       &  $R$ & 1$\times$10 + 1$\times$50 + 3$\times$580 &   1.40-1.45 & 1.1-1.3 \\
8  &  23  55  00 & $-$71  20  00 & 309.1 & -45.1 &  $C$ & 1$\times$60 + 1$\times$300 + 3$\times$1080 & 1.33-1.34 & 1.2-1.3 \\
   &             &               &       &       &  $R$ & 1$\times$10 + 1$\times$50 + 3$\times$580 &   1.35-1.38 & 0.9-1.0 \\
9  &   0  04  02 & $-$70  59  48 & 308.4 & -45.6 &  $C$ & 1$\times$60 + 1$\times$300 + 3$\times$1080 & 1.34-1.38 & 1.0-1.3 \\
   &             &               &       &       &  $R$ & 1$\times$10 + 1$\times$50 + 3$\times$580 &   1.32-1.33 & 1.2-1.4 \\
10 &  23  41  16 & $-$73  08  22 & 309.8 & -43.0 &  $C$ & 1$\times$60 + 1$\times$300 + 3$\times$1080 & 1.37-1.41 & 1.5-1.7 \\
   &             &               &       &       &  $R$ & 1$\times$10 + 1$\times$50 + 2$\times$580 &   1.37 & 1.0 \\
11 &   0  43  21 & $-$70  54  00 & 303.9 & -46.2 &  $C$ & 1$\times$60 + 1$\times$300 + 4$\times$1080 & 1.32-1.34 & 1.4-1.6 \\
   &             &               &       &       &  $R$ & 1$\times$10 + 1$\times$50 + 3$\times$580 &   1.40-1.45 & 1.1-1.3 \\\hline
\end{tabular}
\end{table*}

\begin{table*}
\caption{$CT_1$ data of stars in Field\,1.}
\begin{tabular}{@{}lccccccc}\hline
Star & $\alpha$$_{2000}$ &  $\delta$$_{2000}$ & $T_1$ & $\sigma$($T_1$) & $C-T_1$ & $\sigma$$(C-T_1)$ & n \\
     & $^{(h:m:s)}$ & ($\degr$ $\arcmin$ $\arcsec$) & (mag) & (mag) & (mag) & (mag)  \\\hline
-    &   -     &   -     &  -    &  -    &   -   &   -     \\
    124  &22:32:29.120 &-67:23:03.58  & 18.130   & 0.003  &  1.037  &  0.006 &  5\\
    125  &22:33:54.381 &-67:23:13.03  & 15.868   & 0.003  &  1.338  &  0.004 &  5\\
    126  &22:33:32.216 &-67:23:10.47  & 19.910   & 0.007  &  2.719  &  0.073 &  4\\
-    &   -     &   -     &  -    &  -    &   -   &   -     \\
\hline
\end{tabular}
\end{table*}

\begin{table*}
\caption{Fundamental properties of studied SMC fields.}
\begin{tabular}{@{}lcccc}\hline

Field & $E(B-V)$ & angular & age & [Fe/H] \\
      & (mag)   & distance (deg) & (Gyr) & (dex) \\\hline
1     & 0.024 & 12.8  & $-$  & $-$  \\
2     & 0.026 & 10.9  &  $-$ & $-$ \\
3     & 0.029 & 9.4  & $-$  & $-$ \\
4     & 0.037 & 7.3  & $-$  & $-$ \\
5     & 0.034 & 6.6  & $-$  & $-$ \\
6     & 0.018 & 7.7  & $-$  & $-$ \\
7     & 0.027 & 5.7  & 8$\pm$2  &  -1.0$\pm$0.3\\
8     & 0.029 & 4.7  &  8$\pm$2  & -1.0$\pm$0.3 \\
9    & 0.028 & 4.2  &  8$\pm$3 & -1.0$\pm$0.3 \\
10    & 0.027 & 5.2  &  7$\pm$2 & -1.0$\pm$0.3 \\
11    & 0.033 & 2.1  & 5$\pm$3  &  -1.0$\pm$0.3\\
\hline
\end{tabular}
\end{table*}

\clearpage

\begin{figure*}
\includegraphics[width=144mm]{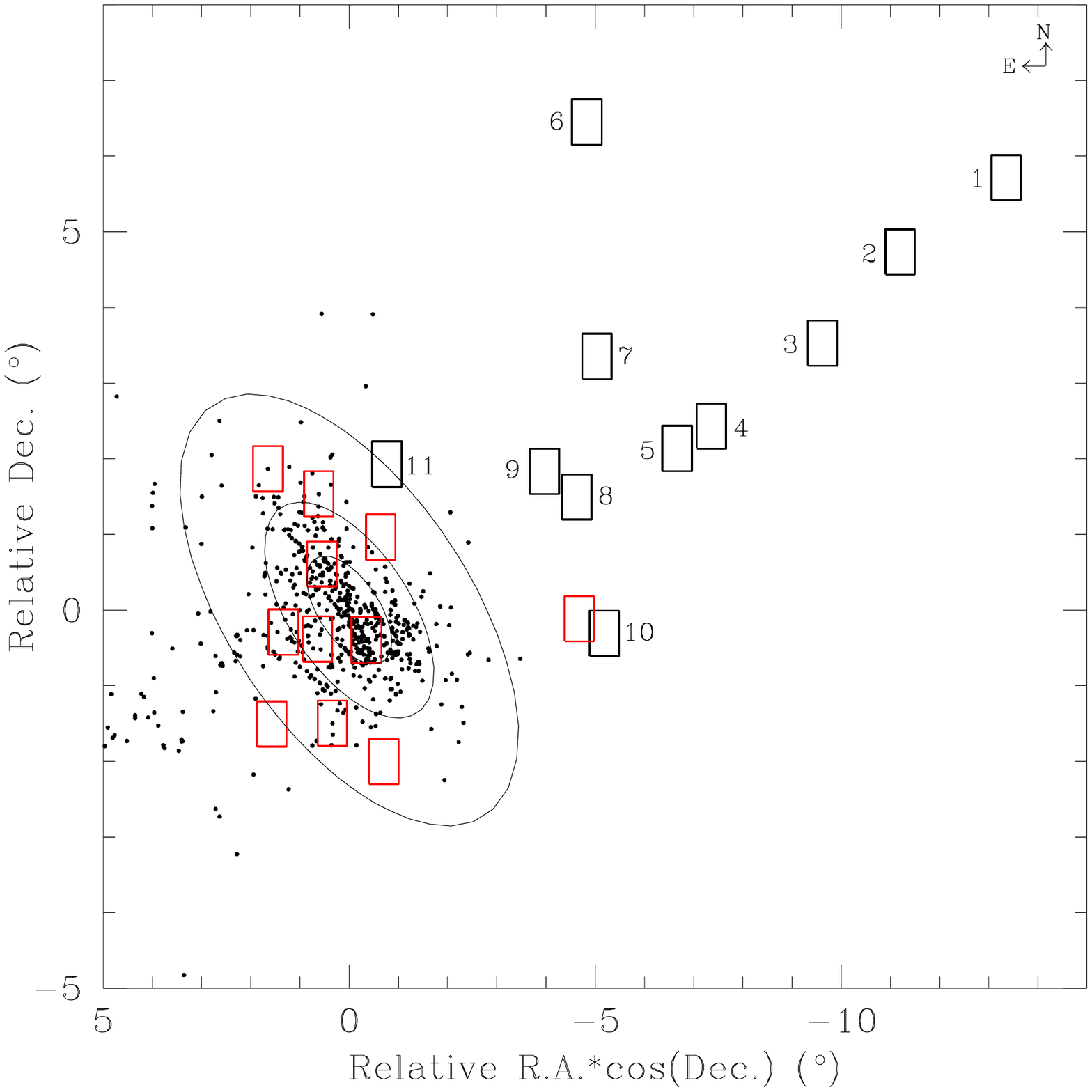}
\caption{Spatial distribution of the presently studied SMC star fields (thick black boxes) and of those
in P12 (thick red boxes). Star clusters catalogued by \citet{betal08} are also drawn (dots) for
comparison purposes. Ellipses with semi-major axes of 1, 2 and 4 deg are overplotted.}
\label{fig1}
\end{figure*}

\begin{figure*}
\includegraphics[width=144mm]{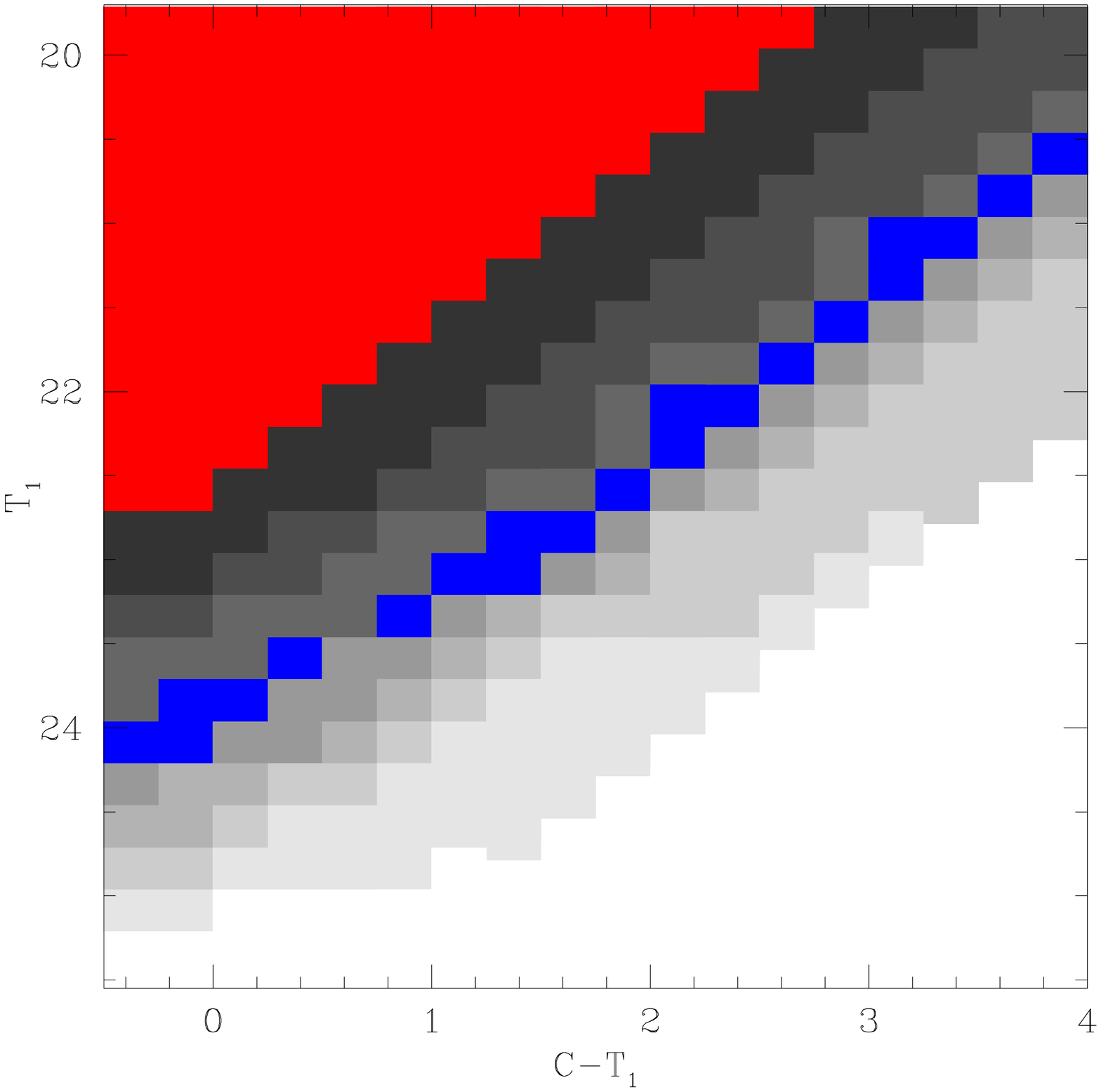}
\caption{Gray-scale completeness fraction for the
crowdest field in our sample (Field\,11). The higher the completeness fraction
the darker the CMD. Red and blue regions represent the 100\% and 50\%
completeness levels, respectively.}
\label{fig2}
\end{figure*}

\begin{figure*}
\includegraphics[width=144mm]{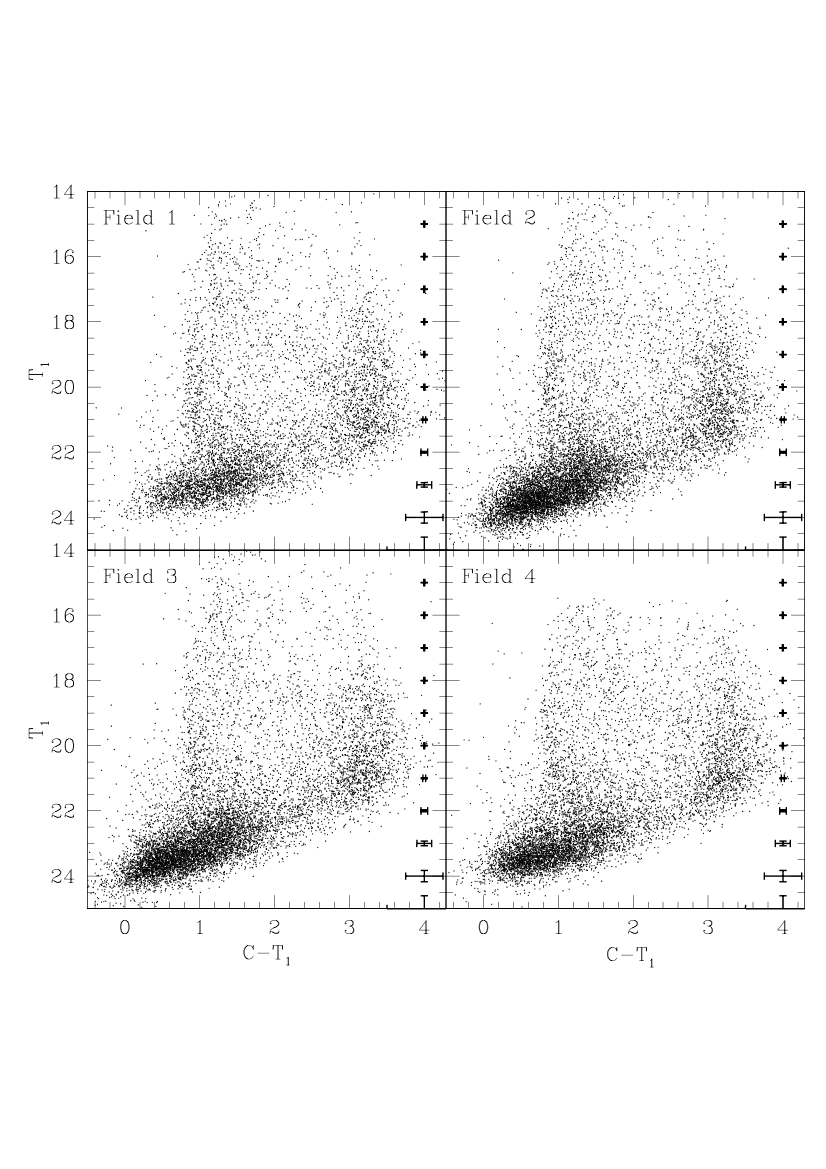}
\caption{CMDs of stars measured in SMC fields.}
\label{fig3a}
\end{figure*}

\setcounter{figure}{2}
\begin{figure*}
\includegraphics[width=144mm]{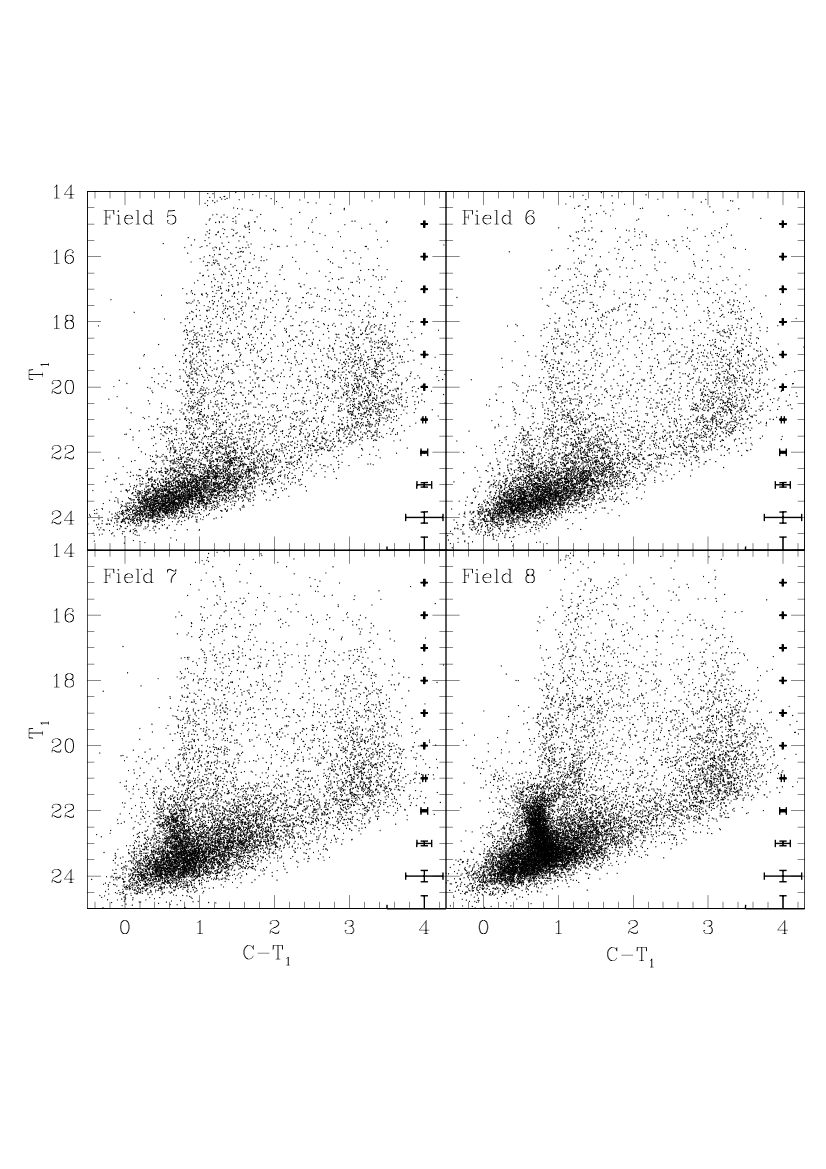}
\caption{continued.}
\label{fig3b}
\end{figure*}

\setcounter{figure}{2}
\begin{figure*}
\includegraphics[width=144mm]{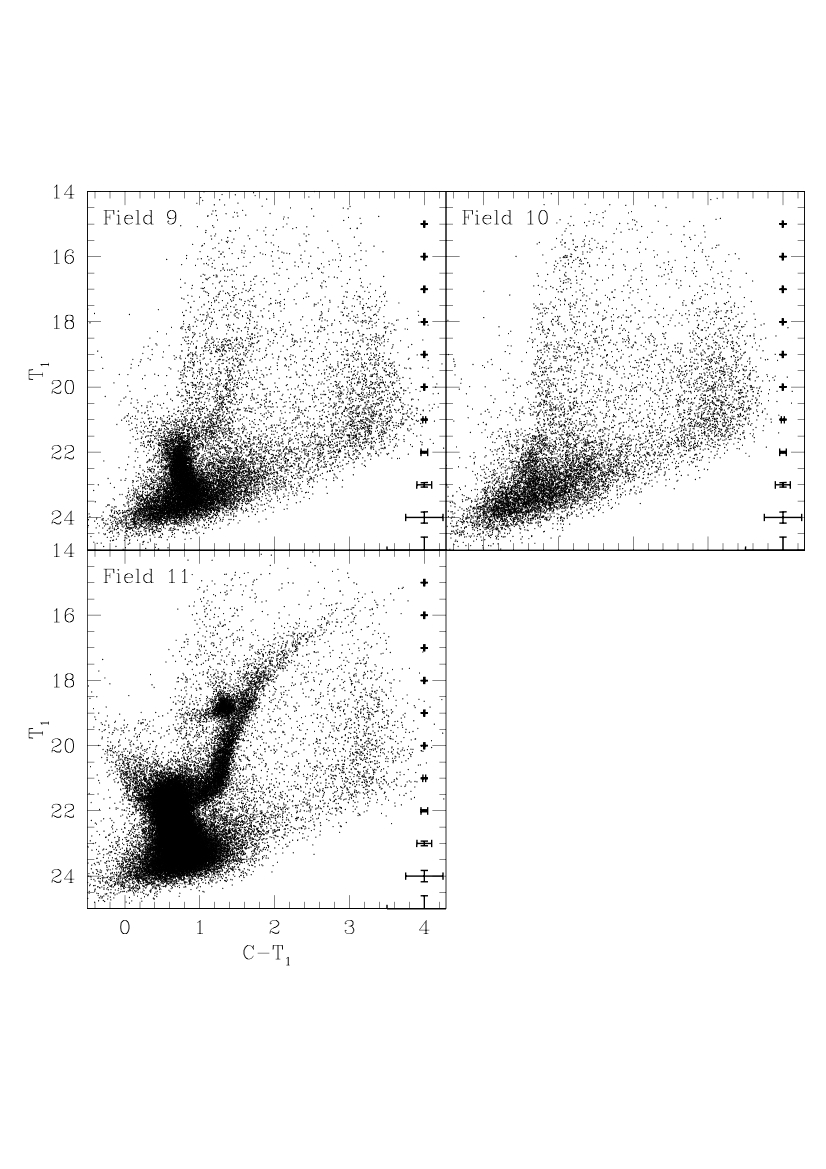}
\caption{continued.}
\label{fig3c}
\end{figure*}

\begin{figure*}
\includegraphics[width=144mm]{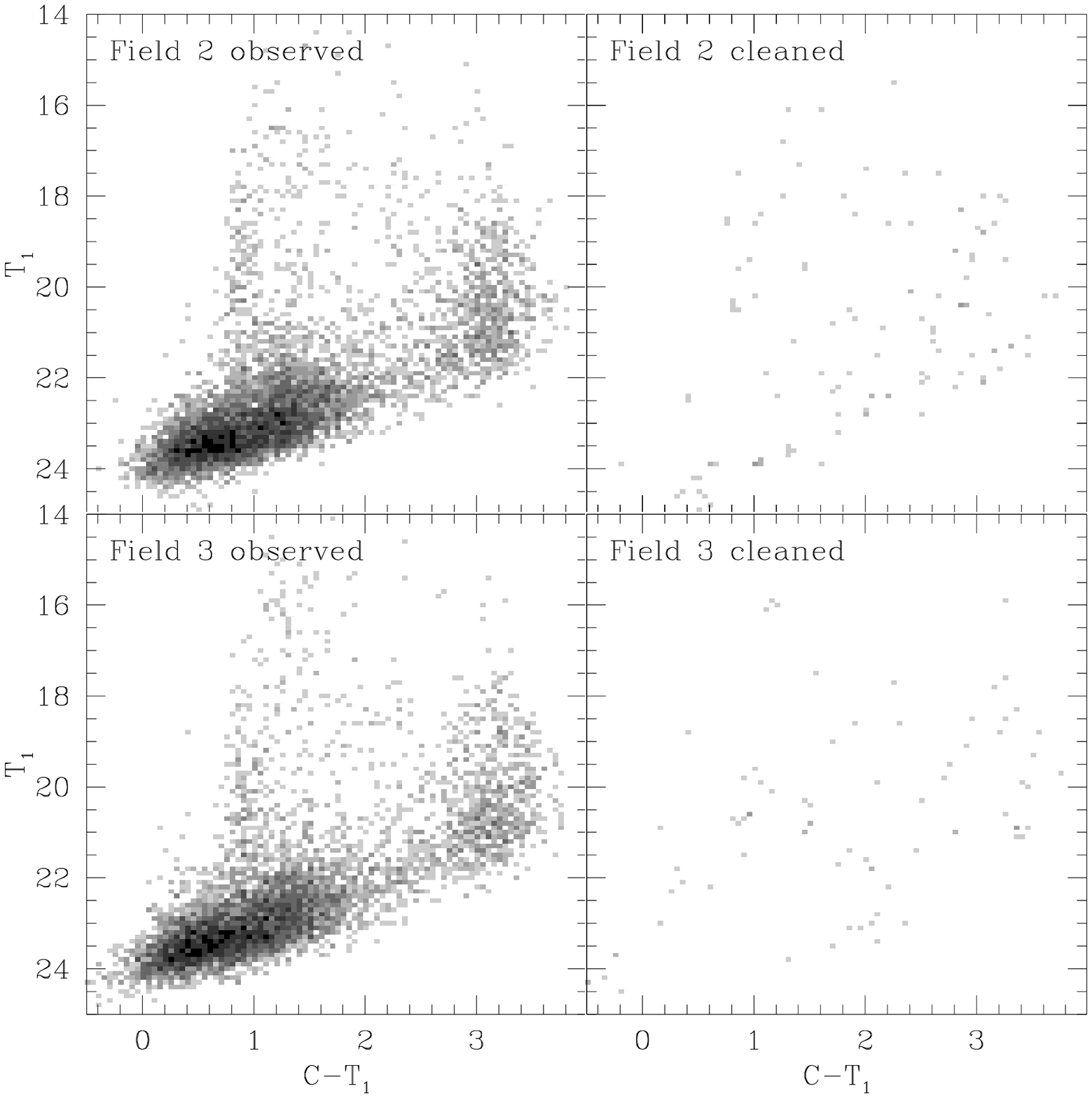}
\caption{Observed and field star decontaminated Hess
diagrams.}
\label{fig4a}
\end{figure*}

\setcounter{figure}{3}
\begin{figure*}
\includegraphics[width=144mm]{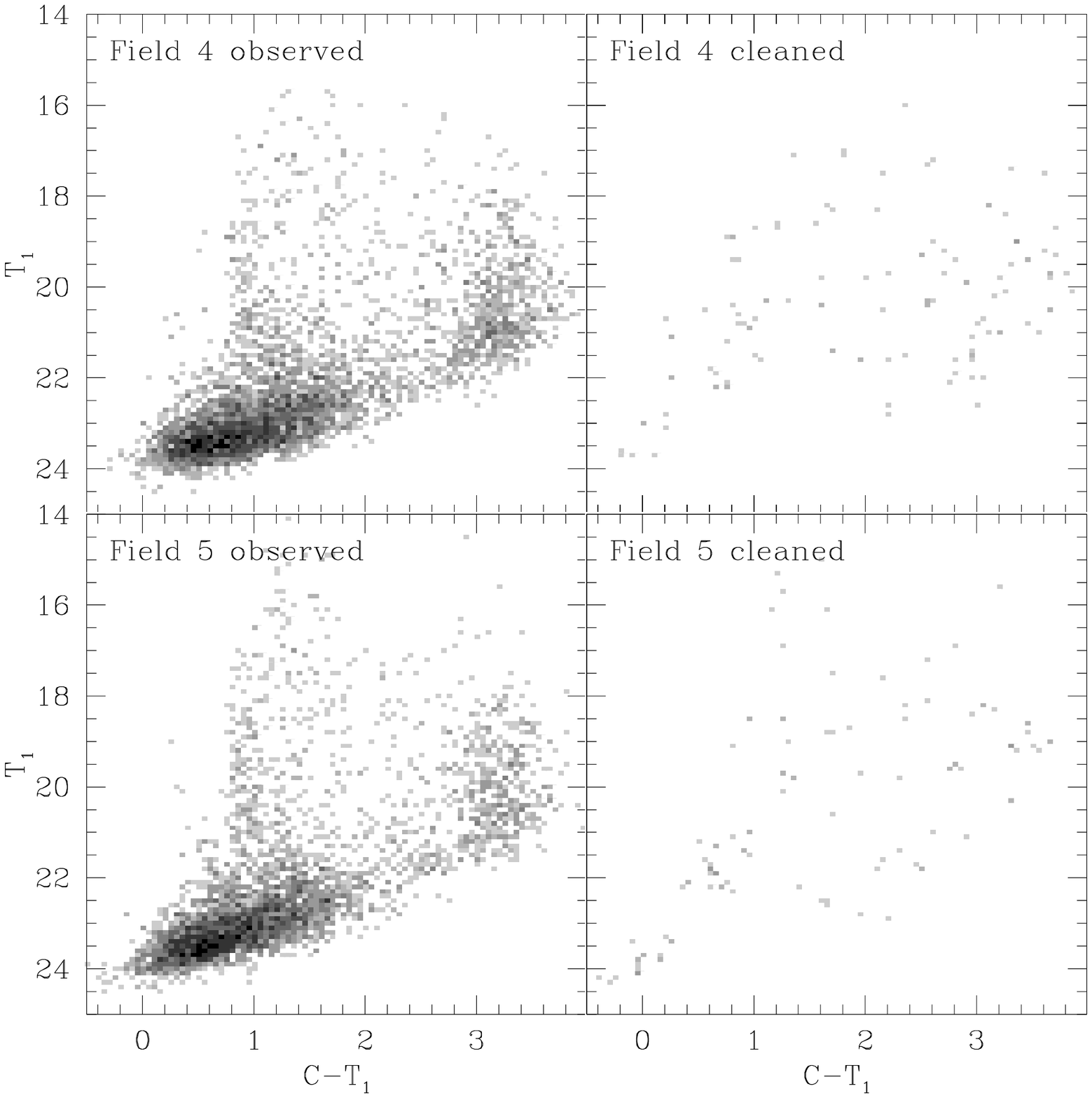}
\caption{continued.}
\label{fig4b}
\end{figure*}

\setcounter{figure}{3}
\begin{figure*}
\includegraphics[width=144mm]{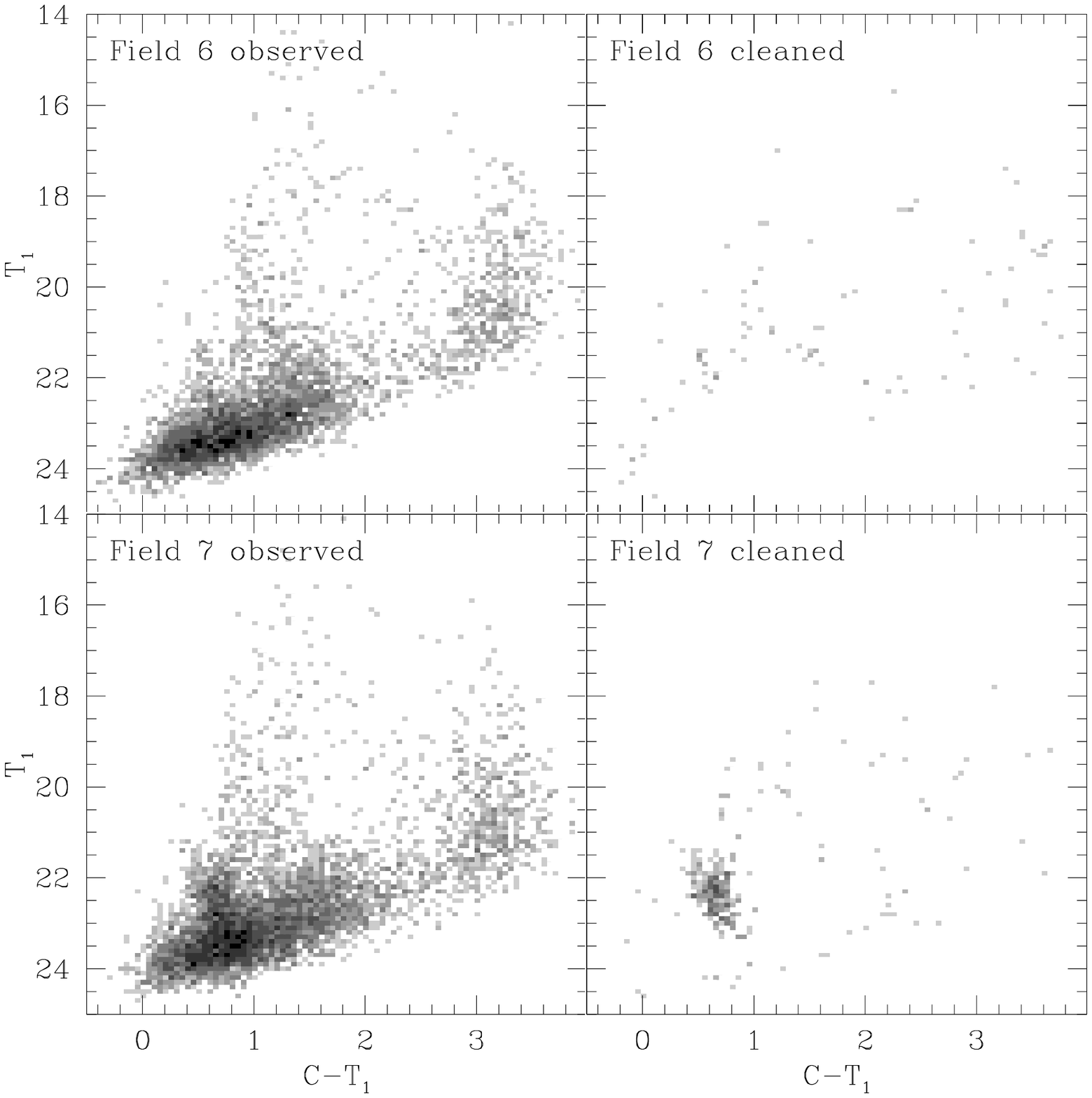}
\caption{continued.}
\label{fig4c}
\end{figure*}

\setcounter{figure}{3}
\begin{figure*}
\includegraphics[width=144mm]{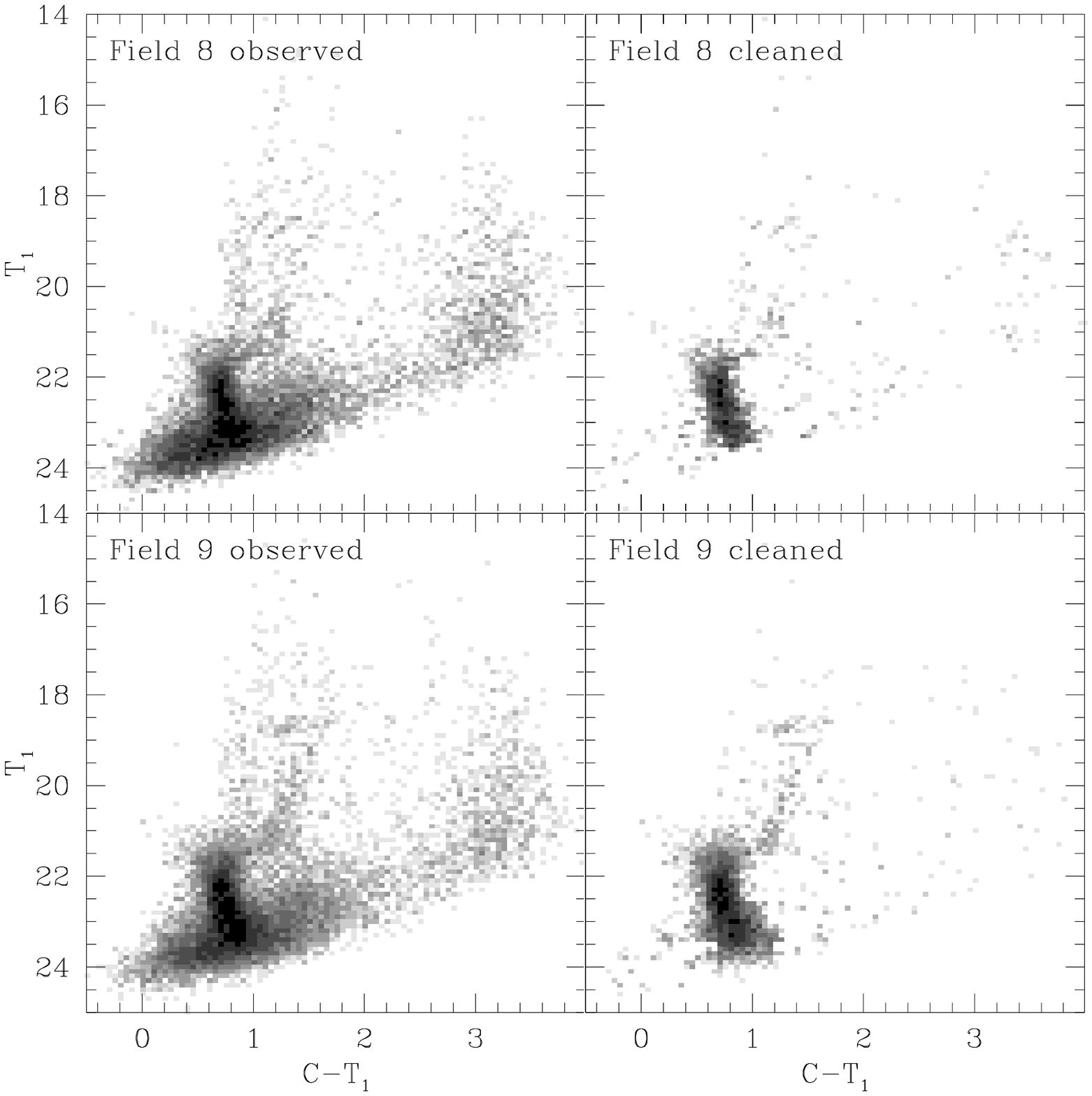}
\caption{continued.}
\label{fig4d}
\end{figure*}

\setcounter{figure}{3}
\begin{figure*}
\includegraphics[width=144mm]{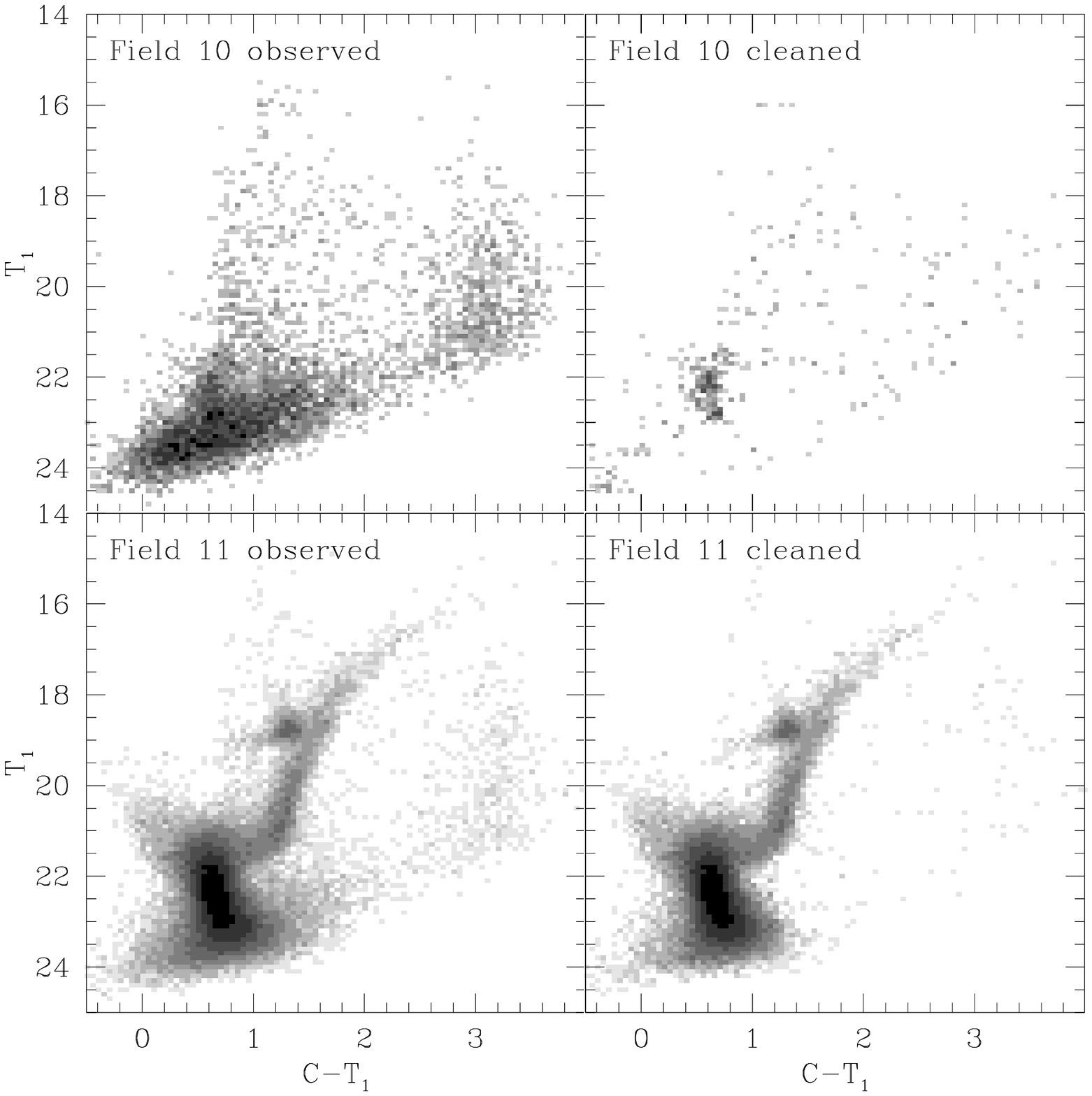}
\caption{continued.}
\label{fig4e}
\end{figure*}

\begin{figure*}
\includegraphics[width=144mm]{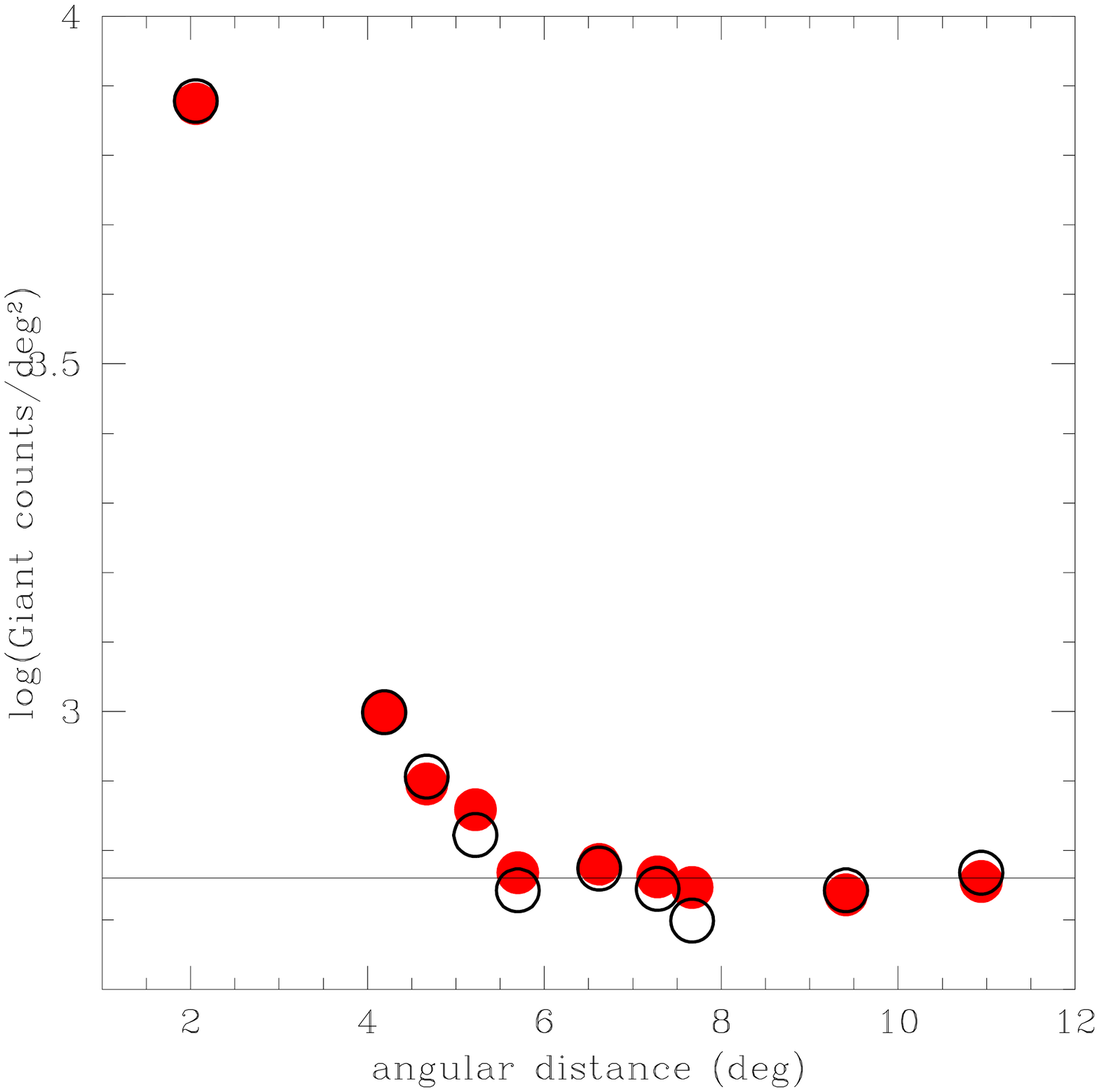}
\caption{Counts of red giant clump stars within the box 
($\Delta$($C-T_1$),$\Delta$($T_1$)) = (0.95$-$1.70, 17.90$-$19.70)
in the cleaned (red circle) and observed (open circle) CMDs of the 
studied SMC fields in terms of their angular distances to the SMC centre. 
The former have beed shifted by a constant for comparison
purposes. The horizontal line represents the counts level for the outermost 
studied field (Field\,1).}
\label{fig5}
\end{figure*}

\begin{figure*}
\includegraphics[width=144mm]{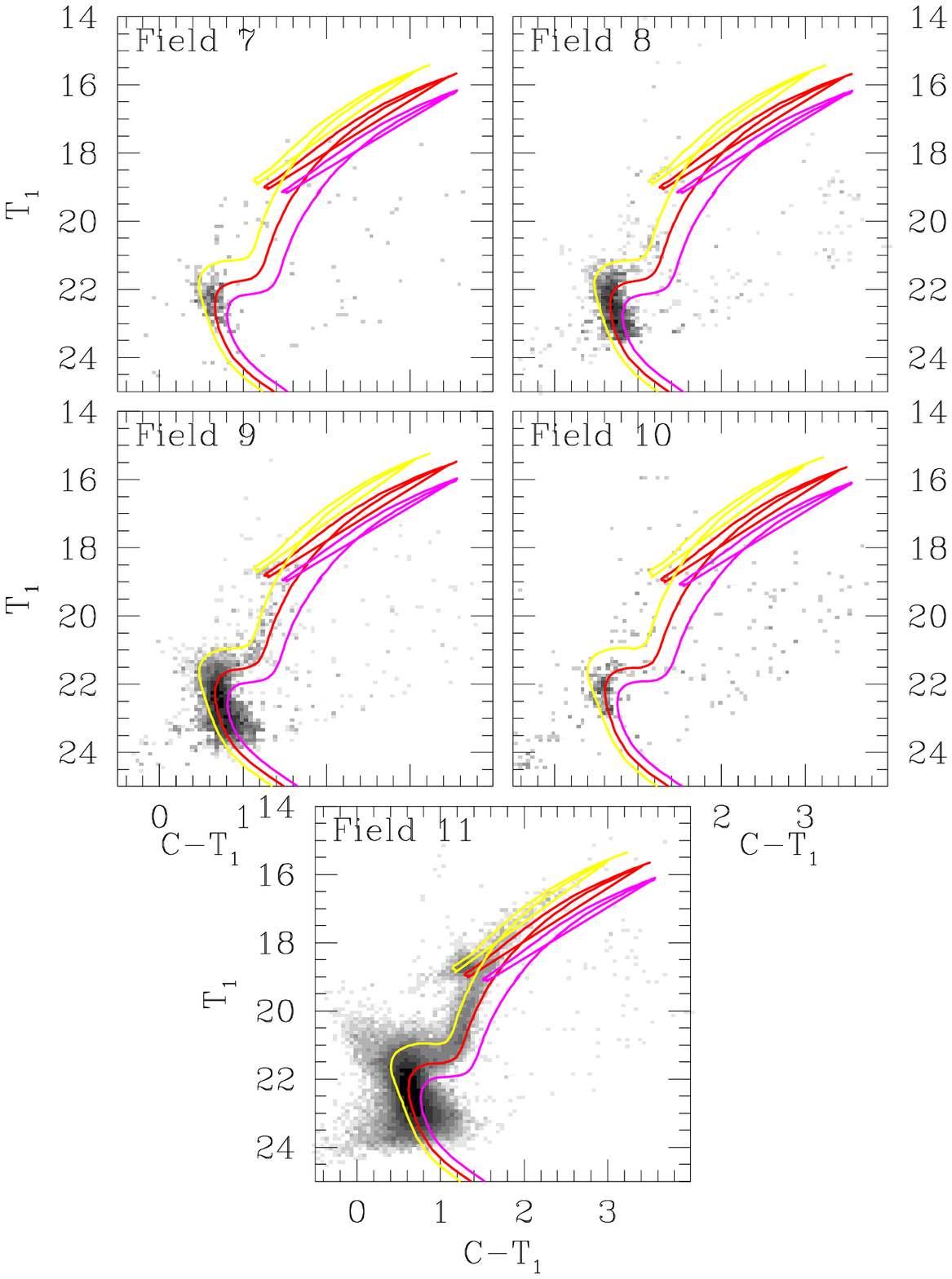}
\caption{Hess diagrams for studied fields with SMC signatures. Three
theoretical isochrones taken from \citet{betal12} were superimposed
as reference of the dominant ages and metallicities: magenta, red and yellow lines
correspond to (age, [Fe/H]) =
(5 Gyr, -1.3 dex), (8 Gyr, -1.0 dex) and (10 Gyr, -0.7 dex), respectively, for
Field\,7, 8 and 9, and  (age, [Fe/H]) =
(4 Gyr, -1.3 dex), (6.3 Gyr, -1.0 dex) and (8 Gyr, -0.7 dex), respectively,
for Field\,10 and 11.}
\label{fig6}
\end{figure*}


\begin{figure*}
\includegraphics[width=144mm]{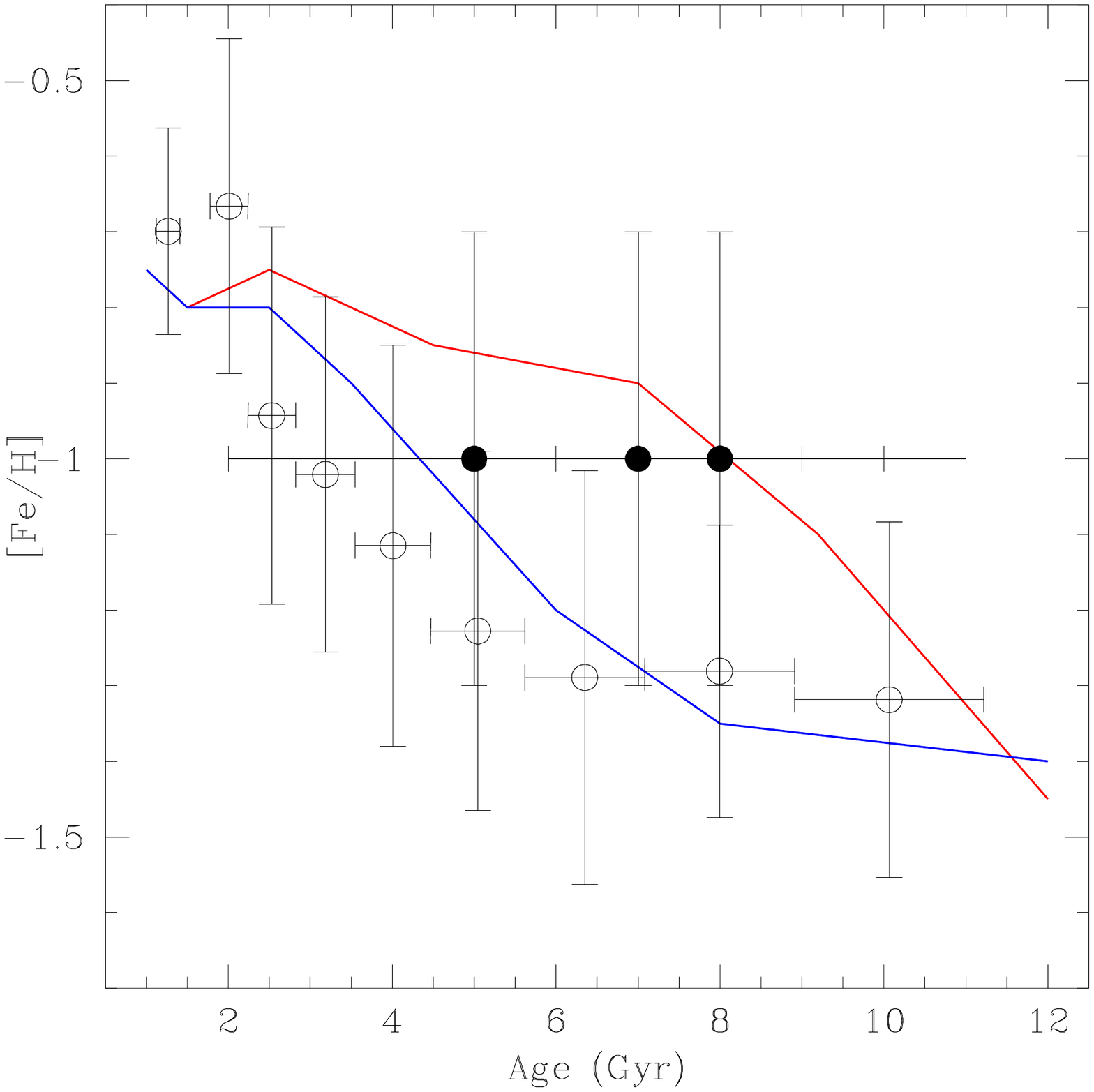}
\caption{AMR for the presently studied SMC star fields (solid circles), with the dispersion
in age and in metallicity represented by errorbars. The mean AMRs by \citet[][red line]{cetal08},
\citet[][blue line]{netal09} and \citet[][open circles with errorbars]{pg13} are overplotted.
}
\label{fig7}
\end{figure*}



\label{lastpage}
\end{document}